# Impact of the electrode material on the performance of light-emitting electrochemical cells

Anton Kirch[1,#], So-Ra Park[1,#], Joan Ràfols-Ribé[1,2], Johannes A. Kassel[3], Xiaoying Zhang[1], Shi Tang[1,2], Christian Larsen[1,2] and Ludvig Edman[1,2,4]*

[1] The Organic Photonics and Electronics Group, Department of Physics, Umeå University, SE-90187 Umeå, Sweden

[2] LunaLEC AB, Umeå University, SE-90187 Umeå, Sweden

[3] Max Planck Institute for the Physics of Complex Systems, Nöthnitzer Straße 38, 01187 Dresden, Germany

[4] Wallenberg Initiative Materials Science for Sustainability, Department of Physics, Umeå University, SE-90187 Umeå, Sweden

[#] These authors contributed equally.

*E-mail: ludvig.edman@umu.se

# Keywords

# Abstract

Light-emitting electrochemical cells (LECs) are promising candidates for fully solution-processed lighting applications because they can comprise a single active-material layer and air-stable electrodes. While their performance is often claimed to be independent of the electrode material selection due to the in-situ formation of electric double layers (EDLs), we demonstrate conceptually and experimentally that this understanding needs to be modified. Specifically, the exciton generation zone is observed to be affected by the electrode work function. We rationalize this finding by proposing that the ion concentration in the injection-facilitating EDLs depends on the offset between the electrode work function and the respective semiconductor orbital, which in turn influences the number of ions available for electrochemical doping and hence shifts the exciton generation zone. Further, we investigate the effects of the electrode selection on exciton losses to surface plasmon polaritons and discuss the impact of cavity effects on the exciton density. We conclude by showing that the measured electrode-dependent LEC luminance transients can be replicated by an optical model that considers these electrode-dependent effects to calculate the attained light outcoupling of the LEC stack. As such, our findings provide rational design criteria considering the electrode materials, the active-material thickness, and its composition in concert to achieve optimum LEC performance.

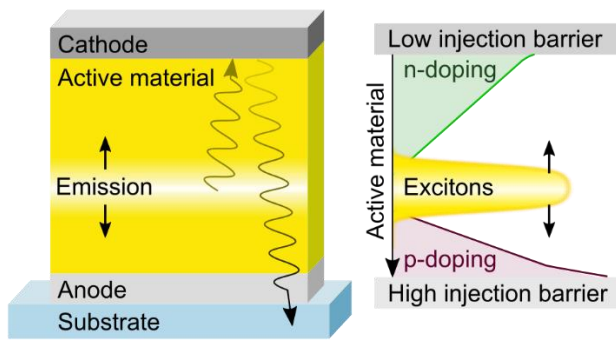

# Introduction

Organic devices can be fabricated by low-energy and resource-efficient solution-based methods using non-toxic materials, which is key to reducing the carbon and waste footprint of optoelectronic products.[1–3] Among others, solution-processed light-emitting diodes (OLEDs),[4] wavelength and oxygen sensors,[5,6] photoluminescent tags (PLTs),[7] photovoltaics (OPVs),[8] and electrochemical transistors (OECTs)[9] have been demonstrated. Light-emitting electrochemical cells (LECs) are particularly appealing in this context, as their entire device structure, i.e. the single active-material layer and the two air-stable electrodes, can be fabricated under ambient conditions[10,11] using non-toxic solvents.[12] From a sustainability perspective, this holds an advantage over advanced and more efficient multilayer p-i-n OLEDs, which are usually fabricated by energy-intense high vacuum processing.[13,14]

In LECs, the single-layer active material (AM) is sandwiched between two electrodes and consists of an emissive organic semiconductor (OSC) and mobile ions. Under applied bias, the mobile ions redistribute and form electric double layers (EDLs) at the electrode/AM interfaces, causing a low resistance for the charge carriers injected into the OSC. The remaining mobile ions drift according to the local electric field and electrically compensate for the space charge generated by the injected electrons and holes, a process called electrochemical (EC) doping. Over time, these n-type and p-type doped regions grow from the cathode and anode, respectively, lowering the transport resistance within the active layer. Electrons and holes meet between the doped regions and generate excitons, which are intended to decay radiatively under the emission of photons. This *exciton generation zone* (EGZ) may also be referred to as *emission zone* or *p-n junction*,[15,16] but since a focal point of this work is on determining and discussing the exciton generation profile, we use the term EGZ.

The in-situ forming doping profiles and thus the EGZ dynamics depend on the LEC driving conditions, the composition of the AM, and the ion and polaron mobility.[15] The dynamic formation of a self-organized doping structure in a single-layer AM stands in contrast to the as-fabricated multi-layer architecture of a p-i-n OLED, where individual, molecularly doped layers enable charge-carrier injection, transport, and recombination at predefined and optimized positions.[17–19]

As for the electrode materials, it is frequently implied that the formation of EDLs renders the LEC performance independent of the electrode material selection.[20–22] This is the reason why LECs are particularly suitable for ambient-air fabrication, where the electrode selection needs to consider its oxidation potential. It has been demonstrated, however, that *extrinsic* degradation related to the electrode material can have a detrimental influence on LEC stability.[23] Notably, if the oxidation (reduction) potential of the positive anode (negative cathode) is positioned at a less positive (negative) potential than the p-type (n-type) doping potential of the OSC, the preferred electrochemical reaction at the anode (cathode) is oxidation (reduction) of the electrode instead of p-type (n-type) doping of the OSC. This undesired scenario leads to electrode degradation and premature device failure.[24] Similarly, it has been shown that other compounds in proximity to the electrode/AM interfaces, e.g. ion transporters, $O_2$ or $H_2O$ impurities,[25] can also cause or be part of electrochemical side reactions.[26,27]

In this study, we show that the choice of the electrode material can have a strong additional *intrinsic* influence on the in-situ forming doping structure in the AM and the physical properties of the optical cavity, two factors that determine the LEC performance. We investigate the impact of the cathode selection (Al, Ag, or Ca) and the AM thickness on the position of the EGZ, on the excitonic coupling to surface plasmon polaritons (SPPs), and the Purcell factor, while keeping a common active-material composition and indium tin oxide (ITO) anode. We conclude by qualitatively replicating the measured electrode-induced luminance changes by an optical simulation.

# Results and discussion

## LEC performance

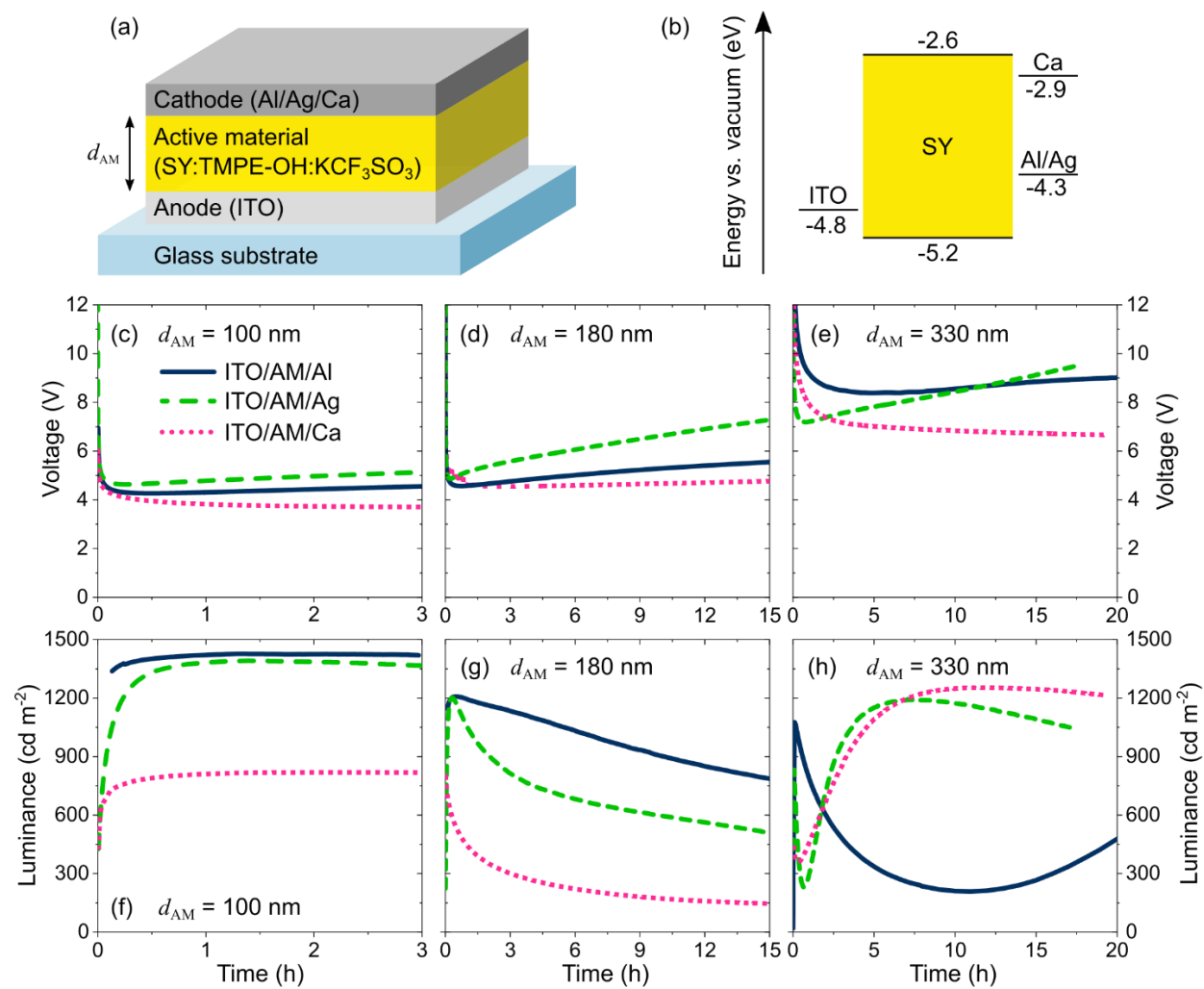

**Figure 1.** (a) LEC structure with the active material (AM) sandwiched between a reflective top cathode and a transparent bottom ITO anode on a glass substrate. The investigated cathode materials are Al, Ag, and Ca, while the investigated AM thicknesses ($d_{AM}$) are 100, 180, and 330 nm. The encapsulation barrier is omitted for clarity. (b) Energy level diagram showing the work function of the electrodes[23] and the HOMO/LUMO levels of Super Yellow (SY).[28] (c-e) Temporal evolution of (c-e) voltage and (f-h) forward luminance of LECs comprising different cathode materials (see label in c) for different $d_{AM}$ under constant current-density operation at 25 mA cm$^{-2}$.

Figure 1a displays the bottom-emitting LEC structure used throughout this work. It comprises a glass substrate, a transparent indium tin oxide (ITO, thickness = 145 nm) anode, an active material (AM) of thickness $d_{AM}$, and a reflective cathode (Al, Ag, or Ca, thickness = 100 nm). The AM consists of the electroluminescent semiconducting polymer Super Yellow (SY), the ion-transporting compound TMPE-OH, and the salt $KCF_3SO_3$, in an optimized mass ratio of 1:0.1:0.03. Three different thicknesses for the AM are investigated: $d_{AM}$ = 100 nm, 180 nm, and 330 nm. Note that, although the fabrication parameters are kept constant, the thickest $d_{AM}$ is found to vary between 320-340 nm between different samples. The $d_{AM}$ variation of a single AM film is invariably below 5 nm, see Experimental section for details. The device is protected from oxygen- and water-induced degradation by an encapsulation glass that is attached by epoxy glue on top of the reflective cathode (not shown in Figure 1a). The Ca cathode consists of 20 nm of Ca in contact with the AM, and 80 nm of Al to protect Ca from contamination during encapsulation.

Figure 1b presents the work function (WF) of the anode and cathode materials[23] as well as the lowest unoccupied molecular orbital (LUMO) and the highest occupied molecular orbital (HOMO) levels of SY.[28] Note that the energy difference between the WF of the cathode (anode) and the LUMO (HOMO) defines the height of the electron (hole) injection barrier. For the three investigated LEC structures, the electron injection barrier varies between 1.7 eV (for Ag and Al) and 0.3 eV (for Ca), while the hole injection barrier from the ITO anode remains unchanged at 0.4 eV.

Figures 1c-h show the temporal evolution of the voltage (c-e) and the forward luminance (f-h) of the nine different LECs when driven at a constant current density of 25 mA cm$^{-2}$. For all devices, the driving voltage decreases, and the luminance increases during the initial operation, cf. Supplementary Information (SI), Figure S1, for a close-up of the initial operation. These two LEC-characteristic observations imply that all devices are well functioning, form injection-enabling EDLs, and develop a doping structure that facilitates charge-carrier transport and recombination by in-situ EC doping.[16]

As expected, the minimum voltage increases with increasing $d_{AM}$, because the EGZ, i.e. the most resistive part of the LEC sporting the lowest doping level, widens with increasing $d_{AM}$.[29] Similarly, the time required to reach the minimum voltage is also increasing with $d_{AM}$ (note the different time scales in Figures 1c-f), as the ions building doped layers need to migrate longer distances.

We further find that the time required to reach the minimum voltage is dependent on the cathode selection, which suggests that the magnitude of undesired, conductivity-degrading side reactions depends on the electrode material. In this context, we note that the electrolyte TMPE-OH:$KCF_3SO_3$ has been demonstrated to exhibit a reduction potential in the proximity of the LUMO level of SY. This implies that EC side reactions can take place in parallel with the preferred EC n-type doping of SY. The observation that the Ag-cathode, and to a lesser extent the Al-cathode, devices exhibit a faster increase in voltage with time thus suggests that Ag, and to a lesser extent Al, "catalyzes" such a conductivity-degrading side reaction.

More surprisingly, the cathode selection and $d_{AM}$ have such a strong influence on the luminance and its transient. Since all samples comprise the same AM, these luminance deviations are suspected to originate from the differing optical environments in each device stack. In the following, we will therefore investigate the impact of the cathode selection and $d_{AM}$ on the EGZ, the coupling of excitons to surface plasmon polaritons, and eventually the properties of the optical cavity.

# Center of the exciton generation zone (CEG)

As for all thin-film electroluminescent (EL) devices, the position of the EGZ can strongly influence the light generation and outcoupling efficiency in LECs, which eventually determines the perceived luminance[18,30]. Since the formation of the doping structure is a dynamic process in LECs, the temporal evolution of the EGZ is crucial information to understand the luminance transients encountered in Figure 1f-h. We determine the center of the exciton generation zone (CEG) of the nine different LECs by measuring their angle-resolved emission spectra with a spectro-goniometer. These data are compared to the angle-dependent emission spectra generated by an optical model of the LEC using the commercial software *Setfos*. The model assumes a Gaussian-shaped exciton generation profile $G(x)$ in (m$^{-3}$s$^{-1}$) with center position CEG and full width at half maximum FWHM$_{EG}$ as fitting parameters, which are optimized via an estimation algorithm. The best-fitting estimator CEG, i.e. the CEG that minimizes the mean squared error between measurement and simulation, discloses where the center of the EGZ is located in the AM, see Experimental section and References [15,31] for details.

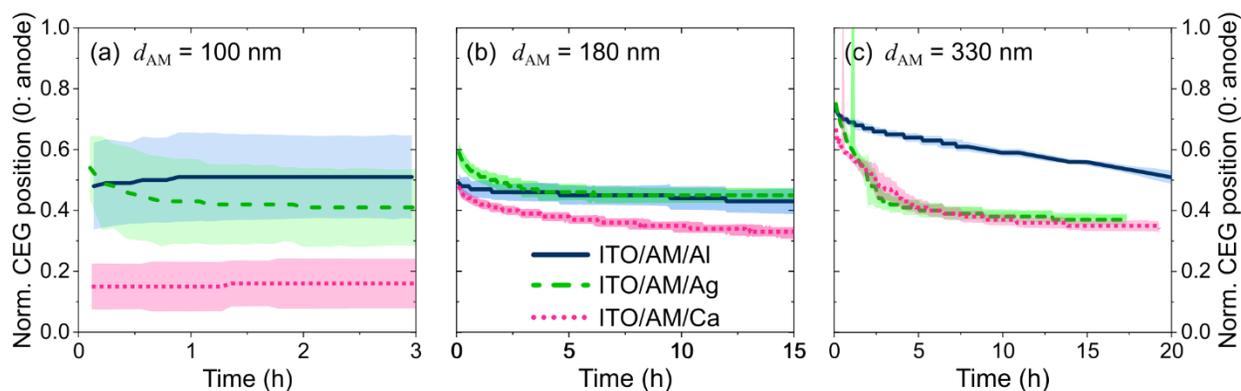

**Figure 2.** Temporal evolution of the center of the exciton generation zone (CEG) in LECs with different cathode materials (see legend in b) and the different values of $d_{AM}$ (a-c). The CEG position is normalized to $d_{AM}$ with 0.0 denoting the anodic and 1.0 the cathodic interface. The lines indicate the best fit between the simulated and measured angle-dependent emission spectra. The shaded areas represent the confidence intervals of the estimation, see details in the Experimental section.

Figure 2 displays the derived temporal evolution of the relative CEG position normalized to $d_{AM}$, with 0.0 corresponding to the anodic and 1.0 to the cathodic interface. The shaded areas indicate the confidence intervals of the estimation algorithm, see Experimental section for details. Since cavity effects are less significant for films that are much thinner than the SY emission wavelength (peak emission at about 550 nm), i.e. far from the wavelength interference condition, the confidence intervals are larger for thinner devices.

Figure 2a shows that the CEGs for the Al and Ag cathodes ($d_{AM}$ = 100 nm) stabilize within about 1 hour close to the center of the AM. The CEG for the Ca cathode, however, is significantly displaced toward the anode (dotted magenta line). A similar trend is observed for $d_{AM}$ = 180 nm in Figure 2b, although the relative difference between the CEGs is smaller. For the two thicker configurations in Figures 2b-c, the EGZs form initially closer to the cathode and subsequently migrate toward the center, which indicates that the anion mobility is smaller than the cation mobility.[15] Notably, the stabilization process of the CEG takes longer for thicker films, which is why the time scales (x-axes) are adjusted for different $d_{AM}$. As introduced above, the Ag and, to a lesser extent, the Al cathodes catalyze a side reaction at the cathode/AM interface. Over time, we suspect this degradation will induce additional errors in the obtained CEG transients, which are hard to quantify. Therefore, the long-term CEG assessment, especially for the thick films, may be corrupted.

A key question is now why the thin Ca features a displaced CEG compared to the corresponding Al and Ag devices. As the CEG is located between p- and n-doped regions in the AM, the observation indicates that the doping profiles are shifted when the cathode material is changed from Ca to Al or Ag. In this context, it is important to note that the electron injection barrier is much smaller with Ca as the cathode (0.3 eV) than with Al or Ag (1.7 eV), cf. Figure 1b. In a running device, this reduced potential drop across the EDL results in a lower cation concentration in the cathodic EDL[32] and increases the number of cations that remain available for EC n-doping. We propose that these excess ions produce a net shift of the doping structure and the CEG toward the anode.

To rationalize this hypothesis, one can estimate the number of ions that are consumed in an EDL by treating the EDL as a parallel-plate capacitor. Assuming a plate (charge) separation of 0.5 nm, one can estimate that about 7 % of all cations available in the AM form the EDL at the AM/Al and AM/Ag interface for $d_{AM}$ = 100 nm, while only about 1 % of them are required in the AM/Ca interface, see SI Section 2 for the detailed estimation. This reduction in consumed cations for the Ca configuration increases the maximum attainable n-doping level. Thus, it decreases the driving voltage under constant-current conditions and causes a shift of the CEG toward the anode. This effect would be more pronounced for a small $d_{AM}$, as the number of ions in the EDL scales with the surface area of the device (unaffected by $d_{AM}$) but the total number of available ions increases with the volume of the AM (linearly with $d_{AM}$). This reasoning can explain both the encountered CEG shift for the ITO/Ca configuration, as well as its lower observed driving voltage, cf. Figure 2.

The impact of this EDL-induced CEG shift depends on the actual number of mobile ions in the AM. If we assume that all salt complexes that are experimentally incorporated into the AM dissociate into mobile ions, i.e. contribute to either the EDL formation or EC doping, and that Al or Ag (Ca) capture merely 7% (1%) of the cations in the cathodic EDLs, 93% (99%) of the cations remain available for EC doping and the overall effect on the CEG and the driving voltage would be minor. However, previous studies propose that a significant share of ions does not contribute to either of the two processes[33,34]. This would increase the ratio of mobile ions consumed by the EDLs and make the doping profiles more asymmetric. We are currently working on quantifying how many ions contribute to EC doping by conductivity measurements. If we can confirm that this number is indeed significantly lower than the experimentally introduced salt density, it would build a strong case for the reasoning above and substantially refine the understanding of LEC physics.

## Surface plasmon polariton (SPP) losses

Beyond their impact on the in-situ forming doping structure and the CEG in LECs, the electrode/AM interface can also impact the exciton density via radiative near-field coupling of excitons to surface plasmon polaritons (SPPs). SPPs travel along the electrode/AM interface and are predominantly excited if the excitonic dipole in the AM is vertically oriented[18,30,35]. The radiative nature of SPP coupling causes an increased effective radiative exciton decay rate $k^*_r(x)$ close to the electrode and therefore depopulates the exciton manifold. While polariton losses in OLEDs are usually contained by an appropriate transport layer design[18] or can be utilized to enhance the device lifetime[36], we will show that they are a major loss mechanism for practical thin-film LECs ($d_{AM} \approx 100$ nm) if the EGZ is not kept at a center position (CEG position ≈ 0.5).

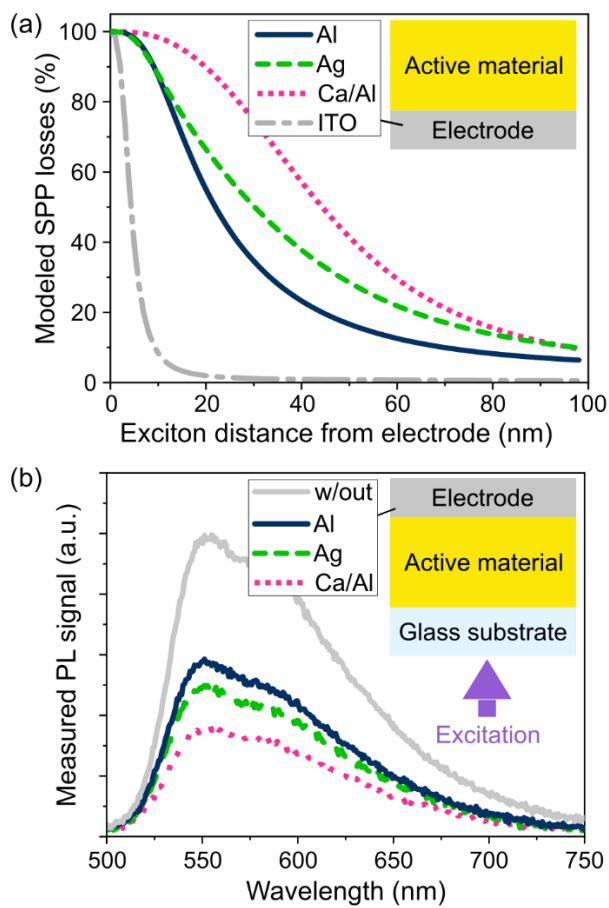

**Figure 3.** (a) The simulated SPP losses of a delta-shaped exciton generation profile as a function of its distance from the electrode surface for the four different electrode materials and an anisotropy coefficient $a = 0.05$. (b) Photoluminescence spectra of a 25 nm thin AM film deposited on a quartz-glass substrate dependent on the top electrode selection as presented in the inset.

To quantify the exciton losses to SPPs and their dependence on the electrode material selection, we use the same optical simulation as above for a slightly adapted stack model. It merely comprises an infinite AM layer on top of a 100 nm thick electrode to exclude further cavity influences (see next section). The electrode material is identified in the inset of Figure 3a. The Ca electrode comprises 20 nm of Ca and 80 nm of Al, as introduced above for the Ca-cathode LECs. The exciton generation profile is modeled as a delta function, using the measured SY anisotropy coefficient $a = 0.05$.[37] The anisotropy coefficient $a$ describes the relative contribution of out-of-plane dipoles to the forward luminance and thus defines the average exciton dipole alignment to the stack normal. It takes values between 0 (horizontal) and 1 (vertical orientation).[38]

Figure 3a presents the simulated ratio of excitons coupling to SPP modes as a function of the spatial separation between the exciton generation profile and the electrode/AM interface for the four employed electrode materials. We find that SPP losses decrease monotonously with increasing exciton-electrode separation for all four electrode materials, which is in line with the established understanding of SPP loss modes for horizontally aligned dipoles[39–41]. According to the simulation, Ca is by far the strongest exciton quencher of the investigated electrode materials, followed by Ag, Al, and ITO.

Figure 3b exemplifies this dependency of SPP-induced exciton losses on the electrode material in a photoluminescence (PL) experiment. Here, we deposit an AM film (same composition as above) with $d_{AM}$ = 25 nm on a quartz glass substrate. On top of that film, one of the indicated cathode materials is evaporated. The AM is excited by UV light ($\lambda_{peak}$ = 450 nm) through the glass substrate, and the resulting PL signal is recorded in an integrating sphere. A sample without an electrode is measured for comparison. The PL intensity is most severely quenched by the Ca electrode, followed by Ag and Al. The highest PL intensity is recorded for the reference sample with no electrode deposited on the AM film. These findings are in line with the model prediction in Figure 3a.

## Purcell factor and exciton density

As for any sandwich EL device, the optical environment influences the exciton dynamics within the AM. When describing the properties of the stack's optical cavity, the complex refractive index ($n$ and $k$ values) of the electrode material influences the reflectivity and phase-shifting properties of the electrode/AM interfaces. The resonance between excitons and reflected photons, together with excitonic coupling to SPPs, produces a local Purcell factor $F(x)$ in the AM[42–44] which alters the natural radiative decay rate $k_{r,0}$, rendering it a position-dependent effective radiative decay rate $k^*_r(x)$.[18,35]

$$k^*_r(x) = F(x) \cdot k_{r,0}.$$

To illustrate the influence of the cathode material on the steady-state exciton density $\rho(x)$, which links to the number of photons generated in the device, we compare the simulated $k^*_r(x)$ for the investigated LECs. This assessment is based on a transfer-matrix algorithm that can be solved in *Setfos*. It takes the optical properties of the cavity and the exciton generation profile $G(x)$ to calculate $k^*_r(x)$ and $\rho(x)$. For this discussion, we assume a common, centered second-order super-Gaussian $G(x)$ with $\mathrm{FWHM_{EG}} = 0.2 \cdot d_{AM}$. This is a Gaussian function with a squared exponent. It yields a flattened center which was found to be a reasonable estimate for real exciton generation profiles[37]. A non-radiative decay rate $k_{nr} = 2 \cdot 10^8$ s$^{-1}$, a natural radiative decay rate $k_{r,0} = 3 \cdot 10^8$ s$^{-1}$, and a drive current density of 25 mA cm$^{-2}$ are used in the simulation.[45] Note that this model is purely optical, i.e. it does not consider Förster-type losses to the electrodes, which non-radiatively depopulate the exciton manifold close to (< 25 nm) the electrodes and thereby exacerbate electrode-induced losses.[46] For additional details on the modeling, see the Experimental section.

Under excitonic steady-state operation and neglecting exciton movement and interaction, the exciton generation $G(x)$ in (m$^{-3}$s$^{-1}$) equals the exciton decay $D(x)$, which is the product of the exciton density $\rho(x)$ in (m$^{-3}$) and the exciton decay rate $k(x)$ in (s$^{-1}$). The latter is specified by its effective radiative and non-radiative components, $k^*_r(x)$ and $k_{nr}$, respectively:

$$G(x) = D(x) = \rho(x) \cdot k(x) = \rho(x) \cdot (k^*_r(x) + k_{nr}).$$

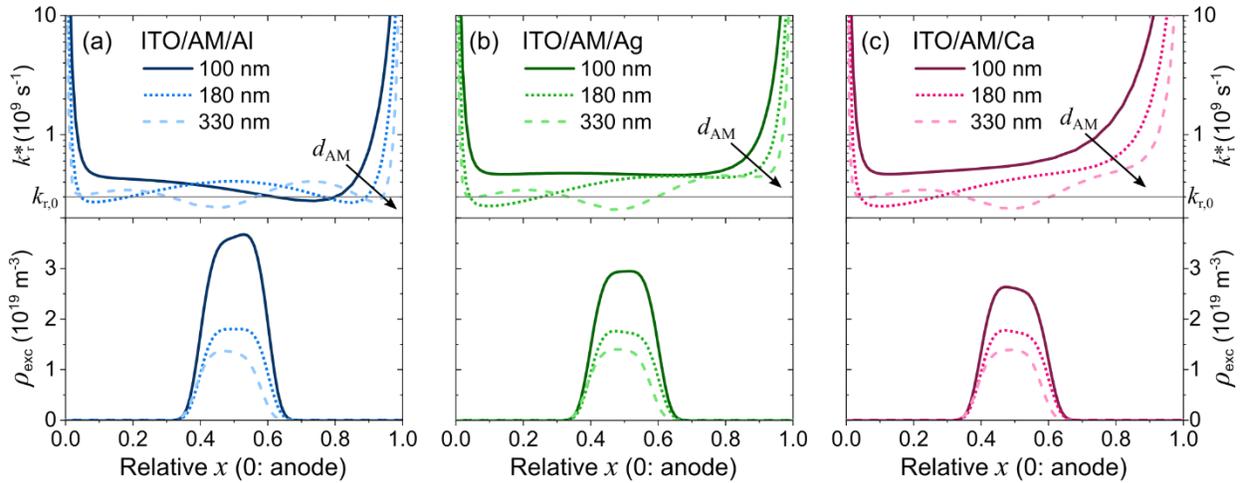

**Figure 4**. Simulated effective radiative exciton decay rate $k^*_r(x)$ (upper panel) and the resulting exciton density $\rho(x)$ (lower panel) as a function of the relative position $x$ in the AM for (a-c) the three different cathode materials and for the three values of $d_{AM}$. The simulation was performed with a common centered second-order super-Gaussian exciton generation profile $G(x)$ at a current density of 25 mA cm$^{-2}$.

Figure 4 presents the simulated values for $k^*_r(x)$ in the upper panel and the resulting $\rho(x)$ in the lower panel as a function of the relative interelectrode position $x$ for the nine different LEC stacks. Note that the anode (ITO) interface is located at $x = 0.0$, the cathode interface at $x = 1.0$, and that the constant relative width of $G(x)$ (FWHM$_{EG}$ = 0.2·$d_{AM}$) results in an absolute decrease of $\rho(x)$ with increasing $d_{AM}$.

The influence range of the electrode-induced SPP losses on $k^*_r(x)$ is clearly visible in the upper panel of Figure 4. It is markedly increased for the Ca cathode, decreases for Ag and Al, and is smallest for ITO, in line with Figure 3. The increasing $k^*_r(x)$ results in a decreasing exciton density when going from Al, over Ag, to Ca for $d_{AM}$ = 100 nm, cf. lower panel in Figure 4 (a to c), and a tilt of $\rho(x)$ corresponding to the respective slope of $k^*_r(x)$ around $x \approx 0.5$. It is important to note that the investigated AM features predominantly horizontally aligned dipoles ($a$ = 0.05). Here, the in-plane wave vector contributions are small and the coupling to SPP modes is only moderate[18,46,47]. For the case of isotropic or even vertical emitter dipole orientation, the impact of SPP losses on $k^*_r(x)$ is even more significant, as exemplified in the SI, Figure S2, and References [48,49].

With increasing $d_{AM}$, interference effects, perceivable by the undulating $k^*_r(x)$, dominate the center of the device, as the absolute distance between excitons and electrodes increases and the optical thickness of the device cavity approaches the emission wavelength of SY. While the SPP influence changes significantly between the cathode materials close to the cathode ($x$ close to 1), these undulating interference patterns remain relatively stable for thick films. This is most apparent for $d_{AM}$ = 330 nm, where $k^*_r(x)$ for $x < 0.7$ looks very similar for all investigated cathode materials, cf. upper panel in Figure 4. Hence, the resulting $\rho(x)$ does not significantly differ between the investigated cathode materials for high $d_{AM}$, cf. lower panel in Figure 4.

We can conclude that thick devices, which are more fit for printing applications, are less susceptible to the choice of the electrode material if the EGZ is located well away from an electrode. Since the SPP losses close to a cathode/AM interface induce the biggest deviations in $k^*_r(x)$ between the investigated cathode materials, it is instead the interference pattern that determines the preferred location for the EGZ. The thinner the device, the stronger the impact of the electrode-dependent coupling to SPP modes. For thin films ($d_{AM}$ = 100 nm), it becomes more important to choose a weak quencher, e.g. ITO, Al, or Ag, as an electrode material.

## Luminance modeling

We conclude our investigation by calculating the outcoupled forward luminance of the LEC model in *Setfos* depending on the three investigated cathode materials. Eventually, we compare the simulated to the experimental luminance data in Figure 1f-h. Again, this is a purely optical assessment that does not include exciton-exciton interactions or polaron quenching, i.e. it assumes a constant non-radiative exciton decay $k_{nr} = 2·10^8$ s$^{-1}$ and no Förster energy-transfer rate. This is why we present the simulated luminance values only qualitatively. For this simulation, the exciton generation profile $G(x)$ is assumed as a delta function to reduce computational effort, the anisotropy factor is set to $a = 0.05$, and the refractive index of SY is used for the AM, see Experimental section for details.

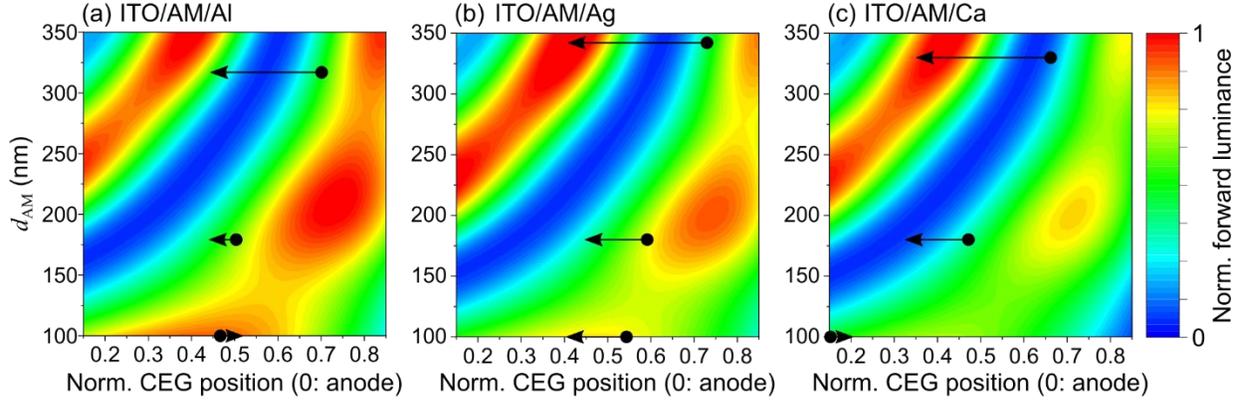

**Figure 5.** Simulated forward luminance as a function of the CEG position (x-axis) and $d_{AM}$ (y-axis) for (a-c) the three investigated cathode materials. The arrows indicate the measured transients of the CEG during LEC operation.

Figure 5 presents the simulated forward luminance of the LEC stack as a function of CEG normalized to $d_{AM}$ (x-axis) and the $d_{AM}$ (y-axis) for the three different cathode materials. These contour plots summarize the impact of the discussed interference and electrode effects and provide straightforward design criteria for optimized LEC luminance. We find that the strongest deviation between the cathode materials can be seen in the bottom and right areas, where the CEG is located closest to the cathode and $d_{AM}$ is small. Moreover, we find that the lowest luminance in this bottom-right area is obtained for the Ca cathode, which is in agreement with Figures 3 and 4, as Ca features the strongest SPP quenching. This finding aligns with the discussion on the impact of the electrode material on $k^*_r(x)$ in the last section: Only for thin devices and close to the electrode, $k^*_r(x)$ is significantly dependent on the electrode material.

To connect these contour plots to the measured luminance data in Figure 1f-h, the derived evolution of the CEGs (c.f. Figure 2) for the nine measured devices are indicated as black arrows. We find that the simulated luminance transients qualitatively reproduce the measured transients in Figure 1f-h. This means that the optical properties of the LEC cavity, depending on the cathode material and $d_{AM}$, and the CEG transients are thus found to be the origin of the differing experimental forward luminance data. This is best illustrated for the thickest devices ($d_{AM} \approx 330$ nm), which start at intermediate luminance, cross a luminance valley, and finally reach almost their luminance maximum, reproducing the encountered luminance undulations in Figure 1h. For the two thinner configurations, the Ca device is expected to operate in an area of lower forward luminance, as excitons couple more heavily to SPP modes, cf. Figure 4c, which aligns with the measurement displayed in Figure 1f-g.

We can conclude that, according to Figure 5, controlling the electrode materials, the CEG, and $d_{AM}$ in concert, based on optical modeling, seems a viable way of finding the conditions of maximum LEC forward luminance.

# Conclusion

In this work, we investigate the influence of the electrode material selection on the performance of a common three-layer LEC stack by combined measurements and simulations. We systematically vary both the cathode material (Al, Ag, and Ca) and the thickness of the active material (between 100 and 330 nm) and find that both parameters heavily influence the LEC performance. Through spectrogoniometer measurements, we study the transients of their exciton generation zones and observe a shift toward the anode for thin films when using a Ca cathode. To explain this observation, we propose that the number of cations forming the cathodic EDL influences the number of ions that remain available for electrochemical doping. In that manner, the exciton generation zone and the device resistance are directly affected by the difference between the cathode WF and the semiconductor LUMO, i.e. the electron injection barrier.

Apart from the impact on the doping structure, we investigate how the electrode material influences exciton losses to surface plasmon polaritons. They are shown to be a dominating and heavily material-dependent loss channel for common thin-film LECs, even if the mean emitter dipole orientation is almost horizontal in our active material. We further discuss the influence of the optical cavity on the effective radiative exciton decay rate and conclude by simulating contour plots of the expected forward luminance for the employed LEC stacks depending on the cathode material. The measured electrode-dependent transients can be replicated qualitatively from these simulations, which shows that the significant electrode-dependent luminance differences are explained by both the cavity's optical properties and the transients of the exciton generation zones.

Thereby, we present evidence that the LEC performance, in contrast to the common conception, is dependent on the electrode material selection and that a rational LEC design should thus collectively consider the electrode material properties, the active-material thickness, and its composition.

# Acknowledgments

The authors acknowledge generous financial support from the Swedish Research Council (2019-02345 and 2021-04778), Kempe Foundations, and the Wallenberg Initiative Materials Science for Sustainability (WISE) funded by the Knut and Alice Wallenberg Foundation (WISE-AP01-D02). L.E. acknowledges financial support from the European Union through an ERC Advanced Grant (ERC, InnovaLEC, 101096650). A.K. acknowledges funding from the European Union (HORIZON MSCA 2023 PF, acronym UNID, grant number 101150699). A.K. thanks Mikael Fredriksson for his support in the workshop.

# Data availability statement

All relevant experimental data and the *Setfos* simulation files are available for download here:

https://doi.org/10.6084/m9.figshare.27248310.v1

# Experimental

## Ink fabrication

The active material comprises a blend of an electroluminescent conjugated polymer, a phenyl-substituted poly(para-phenylenevinylene) copolymer termed "Super Yellow" (SY, Livilux PDY-132, Merck, GER), a hydroxyl end-capped trimethylolpropane ethoxylate (TMPE-OH, $M_n$ = 450 g mol$^{-1}$, Sigma-Aldrich, USA) ion transporter, and a KCF$_3$SO$_3$ (Sigma-Aldrich, USA) salt. The salt (ion transporter) is dried in a vacuum oven at $p < 10^2$ Pa and 190 °C (50 °C) for 12 h before use. The active-material constituents are separately dissolved in cyclohexanone (Sigma-Aldrich, USA) in a concentration of 10-15 g l$^{-1}$ (SY), 10 g l$^{-1}$ (TMPE-OH), and 10 g l$^{-1}$ (KCF$_3$SO$_3$). These master inks are blended in a solute mass ratio of SY:TMPE-OH:KCF$_3$SO$_3$ = 1:0.1:0.03 for the formulation of the active-material ink, which is stirred for ≥ 24 h at 70 °C in a glovebox ([O$_2$] < 1 ppm, [H$_2$O] < 1 ppm).

## Device fabrication

The indium tin oxide (ITO) coated glass substrates (ITO thickness = 145 nm, substrate area = 30×30 mm$^2$, substrate thickness = 0.7 mm, Thin Film Devices, USA) are cleaned by sequential ultrasonication in a detergent (Extran MA 01, Merck, GER), deionized water, acetone (VWR, GER), and isopropanol (VWR, GER) before being dried in an oven at 120 °C for ≥ 12 h. The active-material ink is spin-coated on the ITO substrate at 1000-4500 rpm for 120 s and dried on a hot plate at 70 °C for 1 h. The active-material thickness ($d_{AM}$) is measured with a stylus profilometer (Dektak XT, Bruker, USA). Depending on the SY concentration and spinning parameters, we yield $d_{AM}$ of (100 ± 5) nm, (180 ± 5) nm, and (320/340/330 ± 5) nm for the (Al/Ag/Ca) cathodes, respectively. The Al, Ag, and Ca reflective top cathodes are deposited by thermal evaporation at $p < 2 \times 10^{-4}$ Pa, with a shadow mask defining the cathode area. The spatial overlap between the cathode and the anode defines four 2´2 mm$^2$ LEC pixels on each substrate. The LECs are encapsulated with a cover glass (24×24 mm$^2$, VWR, GER) using a UV-curable epoxy resin (Ossila, UK) and measured under ambient conditions.

## Device characterization

The current-voltage measurements are performed using a computer-controlled source measure unit (SMU 2400, Keithley, USA). The devices are driven by a current density of 25 mA cm$^{-2}$, using a voltage compliance of 21 V. All devices are biased with ITO as the positive anode. The non-polarized, angle-resolved emission spectra and intensity are measured using a custom-built, calibrated spectro-goniometer. The device is placed in a sample holder, which aligns the emission area of the device with the rotation axis of a stepper motor. A fraction of the emitted light is collected by a collimating lens ($\emptyset$ = 7.2 mm, F230 SMA-A, Thorlabs, Germany) positioned 75 mm away from the device. This results in a small and constant solid collection angle ($\Omega$) of 0.007 sr. An optical fiber delivers the collimated light to a CCD-array spectrometer (Flame-S, OceanOptics, USA, linearity >99%, optical resolution FWHM <5 nm). By a rotation of the sample, as controlled by a Python-based virtual instrument, the viewing angle is varied between −80° to +80° in steps of 5° or 10°. The forward luminance is derived from the 0° measurement. A schematic of the setup is depicted in Reference [15], Figure 1. In total, 40 independent devices are measured, cf. SI Section 4, and the presented data is chosen from representative devices within this set.

The photoluminescence (PL) spectra are collected with a commercial PL quantum yield setup (C9920, Hamamatsu Photonics, JP) with the sample placed in an integrating sphere. The sample is excited by a 150 W xenon lamp equipped with a monochromator. The excitation wavelength is set to 450 nm, and the PL signal is recorded by a CCD spectrometer.

## Modeling

All simulations are performed with the commercial software *Setfos* (Version 5.2, Fluxim AG, Switzerland). The model of the device stack comprises the following layers with thicknesses matching the experimental specifications (if not stated otherwise as in Figure 3):

Air (inf.)/reflective top electrode (100 nm of Al, Ag, or Ca/Al)/SY (100, 180, 320, 330, 340 nm)/ITO (145 nm)/Glass (0.75 mm)/Air (inf.).

The CEG in the active material (Figure 2) is derived by finding the absolute minimum mean square error (MSE) between the measured luminous intensity $I_{\theta,\lambda}^{\text{meas}}$ and the simulated luminous intensity $I_{\theta,\lambda}^{\text{sim}}$ for a set of $N_\lambda$ wavelengths $\lambda$ and $N_\theta$ viewing angles $\theta$

$$\text{MSE}\left(I_{\theta,\lambda}^{\text{sim}}(\text{CEG}, \text{FWHM}_{\text{EG}})\right) = \frac{1}{N_\theta}\sum_\theta \frac{1}{N_\lambda}\sum_\lambda \left[I_{\theta,\lambda}^{\text{meas}} - I_{\theta,\lambda}^{\text{sim}}(\text{CEG}, \text{FWHM}_{\text{EG}})\right]^2.$$

The optical model uses a predefined set of exciton generation profiles $G(x)$, each represented by a Gaussian distribution of full width at half maximum FWHM$_{\text{EG}}$ peaking at CEG, to generate the simulated spectral output $I_{\theta,\lambda}^{\text{sim}}(\text{CEG}, \text{FWHM}_{\text{EG}})$ at a given $d_{\text{AM}}$. The assessable peak values for $G(x)$ range between 0.05 and 0.95 (corresponding to the relative interelectrode distance with $x = 0.0$ corresponding to the anode position and $x = 1.0$ to the cathode position) using a grid step size of 0.01. The assessable FWHM$_{\text{EG}}$ values for $G(x)$ range from 0.01 (minimal step size, effectively emulating a delta distribution), 0.1, 0.2, … in steps of 0.1 to 1.0.

Since a narrow $G(x)$ (FWHM$_{\text{EG}}$ < 10 % of $d_{\text{AM}}$) is mostly found as the best fit (except for the $d_{\text{AM}}$ = 330 nm devices shortly after the turn-on), we use a delta distribution to approximate the outcoupling landscape and SPP analysis (Figures 3 and 5) for reduced computational effort. For the analysis of cavity influences on an extended emission zone (Figure 4), we set $G(x)$ as a second-order super-Gaussian centered at $x = 0.5$ and FWHM$_{\text{EG}}$ = 20% of $d_{\text{AM}}$.

The average dipole orientation is set to $a = 0.05$.[37] The emission spectrum of SY is set equal to the measured PL spectrum of a 17-nm thin film of Super Yellow. We set 0.6 as the photoluminescent quantum yield and 2 ns as the natural exciton lifetime in the film, corresponding to $k_{nr} = 2·10^8$ s$^{-1}$ and $k_{r,0} = 3·10^8$ s$^{-1}$.[15,45] The optical constants ($n$ and $k$ values) of the active film are taken from Reference [50] and the optical constants of the electrode materials are taken from the *Setfos* database for SY.

## Confidence interval calculation

The error bands $\sigma_{tot}^{\pm}(t)$ of CEG($t$) in Figure 2 are calculated as the Euclidean norm of two contributing error dimensions at a given point in time $t$:

$$\sigma_{tot}^{\pm}(t) = \sqrt{\sigma_{CEG}^2(t) + [\sigma_{d_{AM}}^{\pm}(t)]^2}.$$

First, we calculate $\sigma_{CEG}$, the standard error of the best-fitting CEG, from the Covariance matrix of the function MSE($x_1$ = CEG, $x_2$ = FWHM$_{EG}$) at the point of optimized parameters $\zeta$. The respective matrix element (1,1) reads

$$\sigma_{CEG} = \sqrt{2 \cdot MSE_\zeta \cdot \left(\frac{\partial^2 MSE_\zeta}{\partial x_i \, \partial x_j}\right)^{-1}_{1,1}}$$

and quantifies the inverse curvature of the MSE landscape[51,52]. A shallow error landscape is thus translated into a high standard error, while a sharp minimum gives a low error. Here, we interpret every spectro-goniometer sweep collecting one spectrum with $N_\lambda$ wavelength bins for $N_\theta$ angles as one single independent measurement ($N$ = 1). Second, the film thickness $d_{AM}$ used for the simulation is varied by ± 5 nm around the experimentally determined value. The resulting shift for the best-fitting CEG is taken as the error $\sigma_{d_{AM}}$, which may differ in both directions ±:

$$\sigma_{d_{AM}}^{\pm} = CEG(d_{AM}) - CEG(d_{AM} \pm 5 \text{ nm}).$$

# References

(1) McCulloch, I.; Chabinyc, M.; Brabec, C.; Nielsen, C. B.; Watkins, S. E. Sustainability Considerations for Organic Electronic Products. *Nat. Mater.* **2023**, *22* (11), 1304–1310. https://doi.org/10.1038/s41563-023-01579-0.
(2) Ren, J.; Opoku, H.; Tang, S.; Edman, L.; Wang, J. Carbon Dots: A Review with Focus on Sustainability. *Advanced Science n/a* (n/a), 2405472. https://doi.org/10.1002/advs.202405472.
(3) Rajendran Nair, R.; Teuerle, L.; Wolansky, J.; Kleemann, H.; Leo, K. Leaf Electronics: Nature-Based Substrates and Electrodes for Organic Electronic Applications. *arXiv e-prints*. July 1, 2024. https://doi.org/10.48550/arXiv.2407.05637.
(4) Su, R.; Park, S. H.; Ouyang, X.; Ahn, S. I.; McAlpine, M. C. 3D-Printed Flexible Organic Light-Emitting Diode Displays. *Science Advances* **2022**, *8* (1), eabl8798. https://doi.org/10.1126/sciadv.abl8798.
(5) Kirch, A.; Bärschneider, T.; Achenbach, T.; Fries, F.; Gmelch, M.; Werberger, R.; Guhrenz, C.; Tomkevičienė, A.; Benduhn, J.; Eychmüller, A.; Leo, K.; Reineke, S. Accurate Wavelength Tracking by Exciton Spin Mixing. *Advanced Materials* **2022**, *34* (38), 2205015. https://doi.org/10.1002/adma.202205015.

(6) Kantelberg, R.; Achenbach, T.; Kirch, A.; Reineke, S. In-Plane Oxygen Diffusion Measurements in Polymer Films Using Time-Resolved Imaging of Programmable Luminescent Tags. *Sci Rep* **2024**, *14* (1), 5826. https://doi.org/10.1038/s41598-024-56237-5.

(7) Gmelch, M.; Achenbach, T.; Tomkeviciene, A.; Reineke, S. High-Speed and Continuous-Wave Programmable Luminescent Tags Based on Exclusive Room Temperature Phosphorescence (RTP). *Advanced Science* **2021**, *8* (23), 2102104. https://doi.org/10.1002/advs.202102104.

(8) Bihar, E.; Corzo, D.; Hidalgo, T. C.; Rosas-Villalva, D.; Salama, K. N.; Inal, S.; Baran, D. Fully Inkjet-Printed, Ultrathin and Conformable Organic Photovoltaics as Power Source Based on Cross-Linked PEDOT:PSS Electrodes. *Advanced Materials Technologies* **2020**, *5* (8), 2000226. https://doi.org/10.1002/admt.202000226.

(9) Massetti, M.; Zhang, S.; Harikesh, P. C.; Burtscher, B.; Diacci, C.; Simon, D. T.; Liu, X.; Fahlman, M.; Tu, D.; Berggren, M.; Fabiano, S. Fully 3D-Printed Organic Electrochemical Transistors. *npj Flex Electron* **2023**, *7* (1), 1–11. https://doi.org/10.1038/s41528-023-00245-4.

(10) Sandström, A.; Asadpoordarvish, A.; Enevold, J.; Edman, L. Spraying Light : Ambient-Air Fabrication of Large-Area Emissive Devices on Complex-Shaped Surfaces. *Advanced Materials* **2014**, *26* (29), 4975–4980. https://doi.org/10.1002/adma.201401286.

(11) Zimmermann, J.; Schlisske, S.; Held, M.; Tisserant, J.-N.; Porcarelli, L.; Sanchez-Sanchez, A.; Mecerreyes, D.; Hernandez-Sosa, G. Ultrathin Fully Printed Light-Emitting Electrochemical Cells with Arbitrary Designs on Biocompatible Substrates. *Advanced Materials Technologies* **2019**, *4* (3), 1800641. https://doi.org/10.1002/admt.201800641.

(12) Larsen, C.; Lundberg, P.; Tang, S.; Ràfols-Ribé, J.; Sandström, A.; Mattias Lindh, E.; Wang, J.; Edman, L. A Tool for Identifying Green Solvents for Printed Electronics. *Nat Commun* **2021**, *12* (1), 4510. https://doi.org/10.1038/s41467-021-24761-x.

(13) Wang, S.-J.; Kirch, A.; Sawatzki, M.; Achenbach, T.; Kleemann, H.; Reineke, S.; Leo, K. Highly Crystalline Rubrene Light-Emitting Diodes with Epitaxial Growth. *Advanced Functional Materials* **2023**, *33* (14), 2213768. https://doi.org/10.1002/adfm.202213768.

(14) Gorter, H.; Coenen, M. J. J.; Slaats, M. W. L.; Ren, M.; Lu, W.; Kuijpers, C. J.; Groen, W. A. Toward Inkjet Printing of Small Molecule Organic Light Emitting Diodes. *Thin Solid Films* **2013**, *532*, 11–15. https://doi.org/10.1016/j.tsf.2013.01.041.

(15) Ràfols-Ribé, J.; Zhang, X.; Larsen, C.; Lundberg, P.; Lindh, E. M.; Mai, C. T.; Mindemark, J.; Gracia-Espino, E.; Edman, L. Controlling the Emission Zone by Additives for Improved Light-emitting Electrochemical Cells. *Advanced Materials* **2022**, *34* (8). https://doi.org/10.1002/adma.202107849.

(16) Matyba, P.; Maturova, K.; Kemerink, M.; Robinson, N.; Edman, L. The Dynamic Organic P-n Junction. *Nature Materials* **2009**, *8* (8), 672–676. https://doi.org/10.1038/nmat2478.

(17) Kirch, A.; Fischer, A.; Werberger, R.; Aabi Soflaa, S. M.; Maleckaite, K.; Imbrasas, P.; Benduhn, J.; Reineke, S. Simple Strategy to Measure the Contact Resistance between Metals and Doped Organic Films. *Phys. Rev. Applied* **2022**, *18* (3), 034017. https://doi.org/10.1103/PhysRevApplied.18.034017.

(18) Brütting, W.; Frischeisen, J.; Schmidt, T. D.; Scholz, B. J.; Mayr, C. Device Efficiency of Organic Light-Emitting Diodes: Progress by Improved Light Outcoupling. *physica status solidi (a)* **2013**, *210* (1), 44–65. https://doi.org/10.1002/pssa.201228320.

(19) Kirch, A. Charge-Carrier Dynamics in Organic LEDs. PhD Thesis, Technische Universität Dresden, 2023. https://nbn-resolving.org/urn:nbn:de:bsz:14-qucosa2-837714.

(20) Matyba, P.; Yamaguchi, H.; Chhowalla, M.; Robinson, N. D.; Edman, L. Flexible and Metal-Free Light-Emitting Electrochemical Cells Based on Graphene and PEDOT-PSS as the Electrode Materials. *ACS Nano* **2011**, *5* (1), 574–580. https://doi.org/10.1021/nn102704h.

(21) Kanagaraj, S.; Puthanveedu, A.; Choe, Y. Small Molecules in Light-Emitting Electrochemical Cells: Promising Light-Emitting Materials. *Advanced Functional Materials* **2020**, *30* (33), 1907126. https://doi.org/10.1002/adfm.201907126.

(22) Chen, F.-C.; Yang, Y.; Pei, Q. Phosphorescent Light-Emitting Electrochemical Cell. *Applied Physics Letters* **2002**, *81* (22), 4278–4280. https://doi.org/10.1063/1.1525881.

(23) Jin, X.; Sandström, A.; Lindh, E. M.; Yang, W.; Tang, S.; Edman, L. Challenging Conventional Wisdom : Finding High-Performance Electrodes for Light-Emitting Electrochemical Cells. *ACS Applied Materials and Interfaces* **2018**, *10* (39), 33380–33389. https://doi.org/10.1021/acsami.8b13036.

(24) Shin, J. H.; Matyba, P.; Robinson, N. D.; Edman, L. The Influence of Electrodes on the Performance of Light-Emitting Electrochemical Cells. *Electrochimica Acta* **2007**, *52* (23), 6456-6462-. https://doi.org/10.1016/j.electacta.2007.04.068.

(25) Fang, J.; Matyba, P.; Robinson, N. D.; Edman, L. Identifying and Alleviating Electrochemical Side-Reactions in Light-Emitting Electrochemical Cells. *Journal of the American Chemical Society* **2008**, *130*, 4562–4568. https://doi.org/10.1021/ja7113294.

(26) Lin, L.; Yang, W.; Liu, Z.; Li, J.; Ke, S.; Lou, Z.; Hou, Y.; Teng, F.; Hu, Y. Oxygen Reduction Reaction Induced Electrode Effects in Polymer Light-Emitting Electrochemical Cells. *Organic Electronics* **2024**, *128*, 107028. https://doi.org/10.1016/j.orgel.2024.107028.

(27) Asadpoordarvish, A.; Sandström, A.; Tang, S.; Granström, J.; Edman, L. Encapsulating Light-Emitting Electrochemical Cells for Improved Performance. *Applied Physics Letters* **2012**, *100*. https://doi.org/10.1063/1.4714696.

(28) Tang, S.; Mindemark, J.; Araujo, C. M. G.; Brandell, D.; Edman, L. Identifying Key Properties of Electrolytes for Light-Emitting Electrochemical Cells. *Chemistry of Materials* **2014**, *26* (17), 5083–5088. https://doi.org/10.1021/cm5022905.

(29) Lindh, E. M.; Lundberg, P.; Lanz, T.; Edman, L. Optical Analysis of Light-Emitting Electrochemical Cells. *Sci Rep* **2019**, *9* (1), 10433. https://doi.org/10.1038/s41598-019-46860-y.

(30) Fuchs, C.; Will, P.-A.; Wieczorek, M.; Gather, M. C.; Hofmann, S.; Reineke, S.; Leo, K.; Scholz, R. Enhanced Light Emission from Top-Emitting Organic Light-Emitting Diodes by Optimizing Surface Plasmon Polariton Losses. *Phys. Rev. B* **2015**, *92* (24), 245306. https://doi.org/10.1103/PhysRevB.92.245306.

(31) Zhang, X.; Ràfols-Ribé, J.; Mindemark, J.; Tang, S.; Lindh, M.; Gracia-Espino, E.; Larsen, C.; Edman, L. Efficiency Roll-Off in Light-Emitting Electrochemical Cells. *Advanced Materials* **2024**, *36* (15), 2310156. https://doi.org/10.1002/adma.202310156.

(32) Groß, A.; Sakong, S. Modelling the Electric Double Layer at Electrode/Electrolyte Interfaces. *Current Opinion in Electrochemistry* **2019**, *14*, 1–6. https://doi.org/10.1016/j.coelec.2018.09.005.

(33) Katsumata, J.; Osawa, F.; Sato, G.; Sato, A.; Miwa, K.; Ono, S.; Marumoto, K. Investigating the Operation Mechanism of Light-Emitting Electrochemical Cells through Operando Observations of Spin States. *Commun Mater* **2023**, *4* (1), 1–10. https://doi.org/10.1038/s43246-023-00366-3.

(34) van Reenen, S.; Janssen, R. A. J.; Kemerink, M. Dynamic Processes in Sandwich Polymer Light-Emitting Electrochemical Cells. *Advanced Functional Materials* **2012**, *22* (21), 4547–4556. https://doi.org/10.1002/adfm.201200880.

(35) Furno, M.; Meerheim, R.; Hofmann, S.; Lüssem, B.; Leo, K. Efficiency and Rate of Spontaneous Emission in Organic Electroluminescent Devices. *Phys. Rev. B* **2012**, *85* (11), 115205. https://doi.org/10.1103/PhysRevB.85.115205.

(36) Zhao, H.; Arneson, C. E.; Fan, D.; Forrest, S. R. Stable Blue Phosphorescent Organic LEDs That Use Polariton-Enhanced Purcell Effects. *Nature* **2024**, *626* (7998), 300–305. https://doi.org/10.1038/s41586-023-06976-8.

(37) Ràfols-Ribé, J.; Hänisch, C.; Larsen, C.; Reineke, S.; Edman, L. In Situ Determination of the Orientation of the Emissive Dipoles in Light-Emitting Electrochemical Cells. *Advanced Materials Technologies* **2023**, *8* (13), 2202120. https://doi.org/10.1002/admt.202202120.

(38) Hänisch, C.; Lenk, S.; Reineke, S. Refined Setup for Angle-Resolved Photoluminescence Spectroscopy of Thin Films. *Phys. Rev. Appl.* **2020**, *14* (6), 064036. https://doi.org/10.1103/PhysRevApplied.14.064036.

(39) Becker, H.; Burns, S. E.; Friend, R. H. Effect of Metal Films on the Photoluminescence and Electroluminescence of Conjugated Polymers. *Phys. Rev. B* **1997**, *56* (4), 1893–1905. https://doi.org/10.1103/PhysRevB.56.1893.

(40) Chance, R. R.; Prock, A.; Silbey, R. Molecular Fluorescence and Energy Transfer Near Interfaces. In *Advances in Chemical Physics*; John Wiley & Sons, Ltd, 1978; pp 1–65. https://doi.org/10.1002/9780470142561.ch1.

(41) Krummacher, B. C.; Nowy, S.; Frischeisen, J.; Klein, M.; Brütting, W. Efficiency Analysis of Organic Light-Emitting Diodes Based on Optical Simulation. *Organic Electronics* **2009**, *10* (3), 478–485. https://doi.org/10.1016/j.orgel.2009.02.002.

(42) Frischeisen, J. Light Extraction in Organic Light-Emitting Diodes. PhD Thesis, Universität Augsburg, 2011.

(43) Pasupathy, K. R.; Ramachandran, A. V.; Ragul, S.; Barah, D.; Mani, R.; Bairava Ganesh, R.; Nair, D. R.; Dutta, S.; Ray, D. Simulation Strategy for Organic LEDs with Spectral Dissimilarity between Photoluminescence and Electroluminescence Spectra. *Opt Quant Electron* **2024**, *56* (8), 1363. https://doi.org/10.1007/s11082-024-07265-y.

(44) Neyts, K. A. Simulation of Light Emission from Thin-Film Microcavities. *J. Opt. Soc. Am. A* **1998**, *15* (4), 962. https://doi.org/10.1364/JOSAA.15.000962.

(45) Rörich, I.; Schönbein, A.-K.; Kamath Mangalore, D.; Ribeiro, A. H.; Kasparek, C.; Bauer, C.; Irina Crăciun, N.; M. Blom, P. W.; Ramanan, C. Temperature Dependence of the Photo- and Electroluminescence of Poly( p -Phenylene Vinylene) Based Polymers. *Journal of Materials Chemistry C* **2018**, *6* (39), 10569–10579. https://doi.org/10.1039/C8TC01998C.

(46) Nowy, S.; Krummacher, B. C.; Frischeisen, J.; Reinke, N. A.; Brütting, W. Light Extraction and Optical Loss Mechanisms in Organic Light-Emitting Diodes: Influence of the Emitter Quantum Efficiency. *Journal of Applied Physics* **2008**, *104* (12), 123109. https://doi.org/10.1063/1.3043800.

(47) Weber, W. H.; Eagen, C. F. Energy Transfer from an Excited Dye Molecule to the Surface Plasmons of an Adjacent Metal. *Opt. Lett., OL* **1979**, *4* (8), 236–238. https://doi.org/10.1364/OL.4.000236.

(48) Barnes, W. L. Fluorescence near Interfaces: The Role of Photonic Mode Density. *Journal of Modern Optics* **1998**, *45* (4), 661–699. https://doi.org/10.1080/09500349808230614.

(49) Celebi, K.; Heidel, T. D.; Baldo, M. A. Simplified Calculation of Dipole Energy Transport in a Multilayer Stack Using Dyadic Green's Functions. *Opt. Express, OE* **2007**, *15* (4), 1762–1772. https://doi.org/10.1364/OE.15.001762.

(50) Lanz, T.; M. Lindh, E.; Edman, L. On the Asymmetric Evolution of the Optical Properties of a Conjugated Polymer during Electrochemical P- and n-Type Doping. *Journal of Materials Chemistry C* **2017**, *5* (19), 4706–4715. https://doi.org/10.1039/C7TC01022B.

(51) Bevan, A. *Statistical Data Analysis for the Physical Sciences*; Cambridge University Press: New York, UNITED STATES, 2013.

(52) Ljung, L. *System Identification: Theory for the User*, 1st edition.; Pearson: Upper Saddle River, NJ, 1999.

# Supporting Information

# Impact of the electrode material on the performance of light-emitting electrochemical cells

Anton Kirch[1,#], So-Ra Park[1,#], Joan Ràfols-Ribé[1,2], Johannes A. Kassel[3], Xiaoying Zhang[1], Shi Tang[1,2], Christian Larsen[1,2] and Ludvig Edman[1,2,4]*

[1] The Organic Photonics and Electronics Group, Department of Physics, Umeå University, SE-90187 Umeå, Sweden

[2] LunaLEC AB, Umeå University, SE-90187 Umeå, Sweden

[3] Max Planck Institute for the Physics of Complex Systems, Nöthnitzer Straße 38, 01187 Dresden, Germany

[4] Wallenberg Initiative Materials Science for Sustainability, Department of Physics, Umeå University, SE-90187 Umeå, Sweden

[#] These authors contributed equally.

*E-mail: ludvig.edman@umu.se

# 1. Experimental data after LEC turn-on

To illustrate the turn-on behavior of the investigated LECs, we plot the same data as in Figure 1, main manuscript, for the first 1000 s. The turn-on characteristics of a functional device are a decreasing driving voltage and increasing luminance during the initial operation. Note that the devices are driven at a constant current density of 25 mA cm$^{-2}$ while setting a voltage compliance of 21 V.

The forward luminance is derived from the 0° spectro-goniometer measurement. The luminance monitoring was accidentally not always started right after the current supply, as the turn-on monitoring is not the scope of this work.

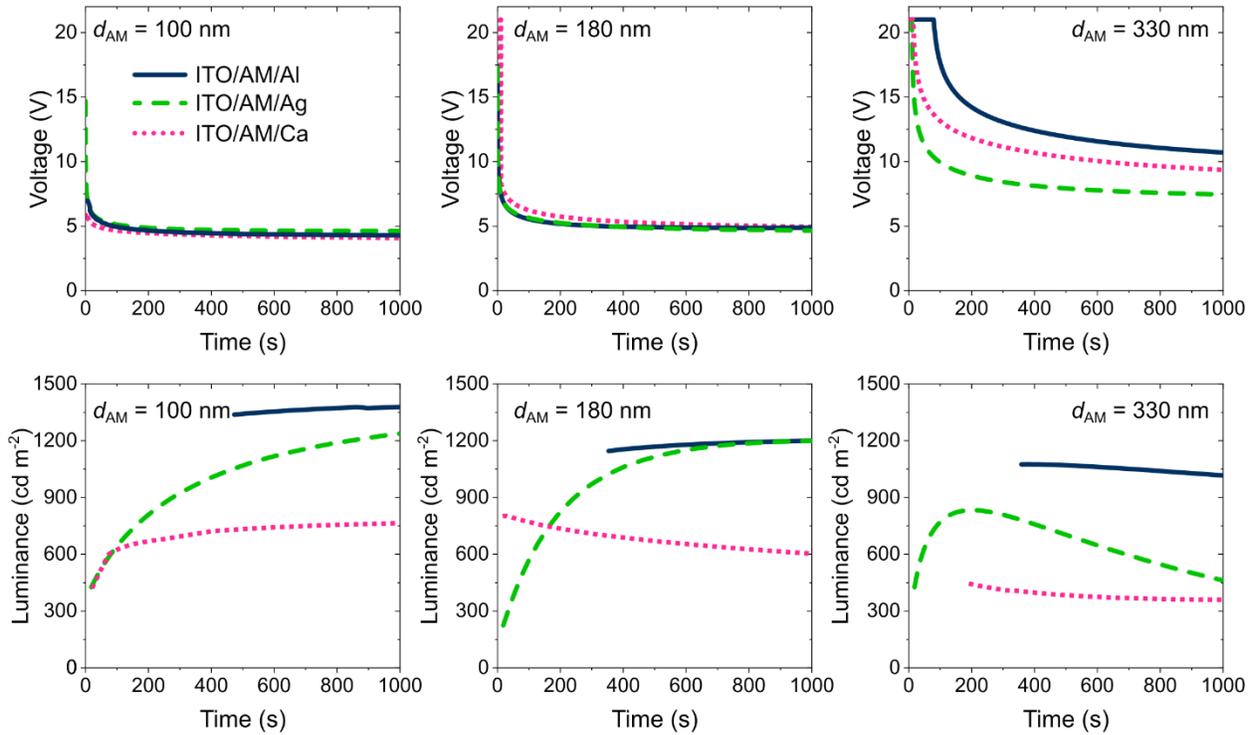

**Figure S1.** LEC characterization (same data as in Figure 1, main manuscript) for the first 1000 s.

## 2. The number of ions in EDLs

The number of non-compensated ions "locked up" in an EDL ($n_{ion,EDL}$) is equal to the net electronic charge on the electrode side of the EDL, and $n_{ion,EDL}$ can be calculated by:

$$C_{EDL} = \frac{Q_{EDL}}{V_{EDL}} = \frac{e \cdot n_{ion,EDL}}{V_{EDL}} = \varepsilon_r \cdot \varepsilon_0 \cdot \frac{A_{EDL}}{d_{EDL}} \quad (1)$$

$$n_{ion,EDL} = \frac{\varepsilon_r \cdot \varepsilon_0 \cdot A_{EDL} \cdot V_{EDL}}{e \cdot d_{EDL}} \quad (2)$$

The variables and parameters in the above equation, with our employed values in parenthesis, are as follows: $C_{EDL}$ is the capacitance of the EDL, $Q_{EDL}$ is the net charge on either side of the EDL, $V_{EDL}$ is the voltage drop over the EDL, $e$ is the elementary charge, $\varepsilon_r$ ($\approx 3$) is the relative permittivity of the active material, $\varepsilon_0$ is the vacuum permittivity, $A_{EDL}$ is the effective cross-section area of the EDL and it is estimated to be approximately equal to the cross-section area of the device ($A_{EDL} \approx A_{AM} = 4 \cdot 10^{-6}$ m$^2$), $d_{EDL}$ ($\approx 0.5 \cdot 10^{-9}$ m) is the effective thickness of the EDL. If we assume that ohmic injection is attained when $V_{EDL}$ exactly compensates the injection barrier at the electrode/active-material (AM) interface, we get that $V_{EDL}$ (Al/AM) = 1.7 V, $V_{EDL}$ (Ag/AM) = 1.7 V, $V_{EDL}$ (Ca/AM) = 0.3 V, and $V_{EDL}$ (ITO/AM) = 0.4 V. By plugging these values into Eq. (2), we find that $n_{ion,EDL}$ (Al/AM) = $n_{ion,EDL}$ (Ag/AM) = $2.3 \cdot 10^{12}$ ions, $n_{ion,EDL}$ (Ca/AM) = $0.4 \cdot 10^{12}$ ions, and that $n_{ion,EDL}$ (ITO/AM) = $0.5 \cdot 10^{12}$ ions.

These numbers can be compared to the *total* number of ions in the active material, which can be calculated with the following equation:

$$n_{ion,AM} = A_{AM} \cdot d_{AM} \cdot \frac{m_{salt}}{m_{AM}} \cdot \rho_{AM} \cdot \frac{1}{M_{salt}} \cdot N_A \cdot 2 \quad (3)$$

Here, $A_{AM}$ (= $4 \cdot 10^{-6}$ m$^2$) is the effective cross-sectional (or emission) area of the LEC, $d_{AM}$ (= 100 or $180 \cdot 10^{-9}$ m) is the thickness of the active material, $\frac{m_{salt}}{m_{AM}}$ (=$\frac{0.03}{1.13}$) is the mass fraction of the KCF$_3$SO$_3$ salt in the active material, $\rho_{AM}$ ($\approx 1 \cdot 10^6$ g m$^{-3}$) is the density of the active material, $M_{salt}$ (= 188.17 g mol$^{-1}$) is the molar mass of the salt, $N_A$ is Avogadro's constant, while the final factor "2" converts the number of salt molecules to the number of ions.

This results in that the active material contains a total of $n_{ion,AM} = 6.8/12.2 \cdot 10^{13}$ ions, divided into $3.4/6.1 \cdot 10^{13}$ K$^+$ cations and $3.4/6.1 \cdot 10^{13}$ CF$_3$SO$_3^-$ anions.

In other words, 1.5/0.8 % of the CF$_3$SO$_3^-$ anions in the active material are "lost" to the anodic EDL at the ITO electrode in all three investigated LECs. In contrast, 6.7/3.7 % of the K$^+$ cations are "lost" at the EDL at the Al and Ag cathodes, while 1.1/0.7 % of the K$^+$ cations are lost at the EDL at the Ca cathode. Since it is the remaining ions, not "lost" to the EDLs that can contribute to the electrochemical doping, this suggests that the maximum attainable p-type doping concentration varies.

# 3. Influence of the emitter orientation on SPP coupling

The impact of SPP coupling on the effective radiative exciton decay rate $k_r^*(x)$ varies significantly with the emissive dipole orientation coefficient $a$. Figure S2 shows $k_r^*(x)$ for the investigated LEC configurations depending on $a$. The case $a = 0.05$ is presented in the main manuscript, Figure 4. A small $a$ indicates preferably horizontal dipole alignment (as for the material Super Yellow used in this study, $a = 0.05$), $a = 1/3$ denotes isotropic (random) dipole orientation, and $a = 1$ means exclusively vertical dipole orientation. The more vertical the mean dipole orientation ($a$ is closer to 1), the more the radiative coupling of excitons to SPP modes is enhanced (e.g. S. Nowy et al., *Journal of Applied Physics* **2008**, *104* (12), 123109).

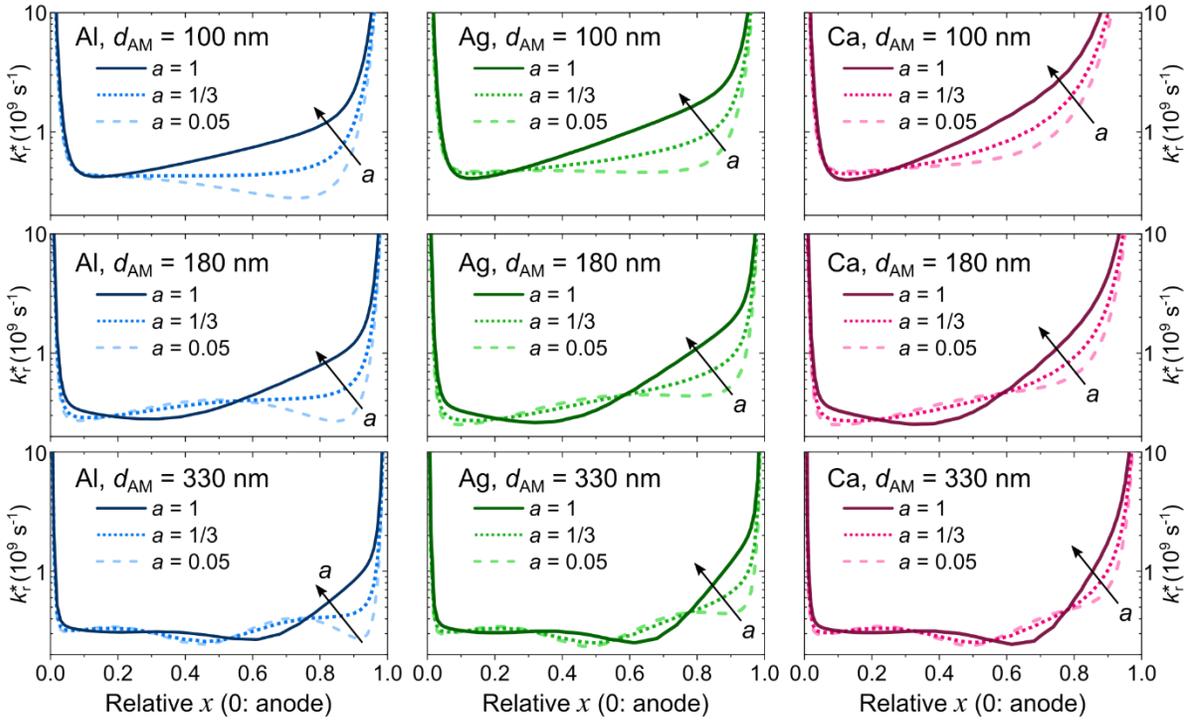

**Figure S2.** The impact of the emitter dipole orientation $a$ on the effective radiative exciton decay rate $k_r^*(x)$ for the investigated LEC configurations. In the main manuscript, the data for $a = 0.05$ is presented.

# 4. Experimental data of all investigated devices

The following Figures S3-S5 show the data of all investigated devices. The data presented in the main manuscript, Figure 1, and in Figure S1 are highlighted with an asterisk (*). If several devices are measured for one LEC configuration, a representative device was chosen or the one that was investigated long enough according to the final data analysis. Note that here the CEG is calculated with a delta-distributed exciton generation profile $G(x)$, which induces errors for the thick-film LECs (perceivable as jumps in Figure S5). Concerning the data selected for the main manuscript (*), the CEG was determined again more accurately using a Gaussian-shaped exciton generation profile with varying thickness FWHM$_{EG}$ and peak position CEG.

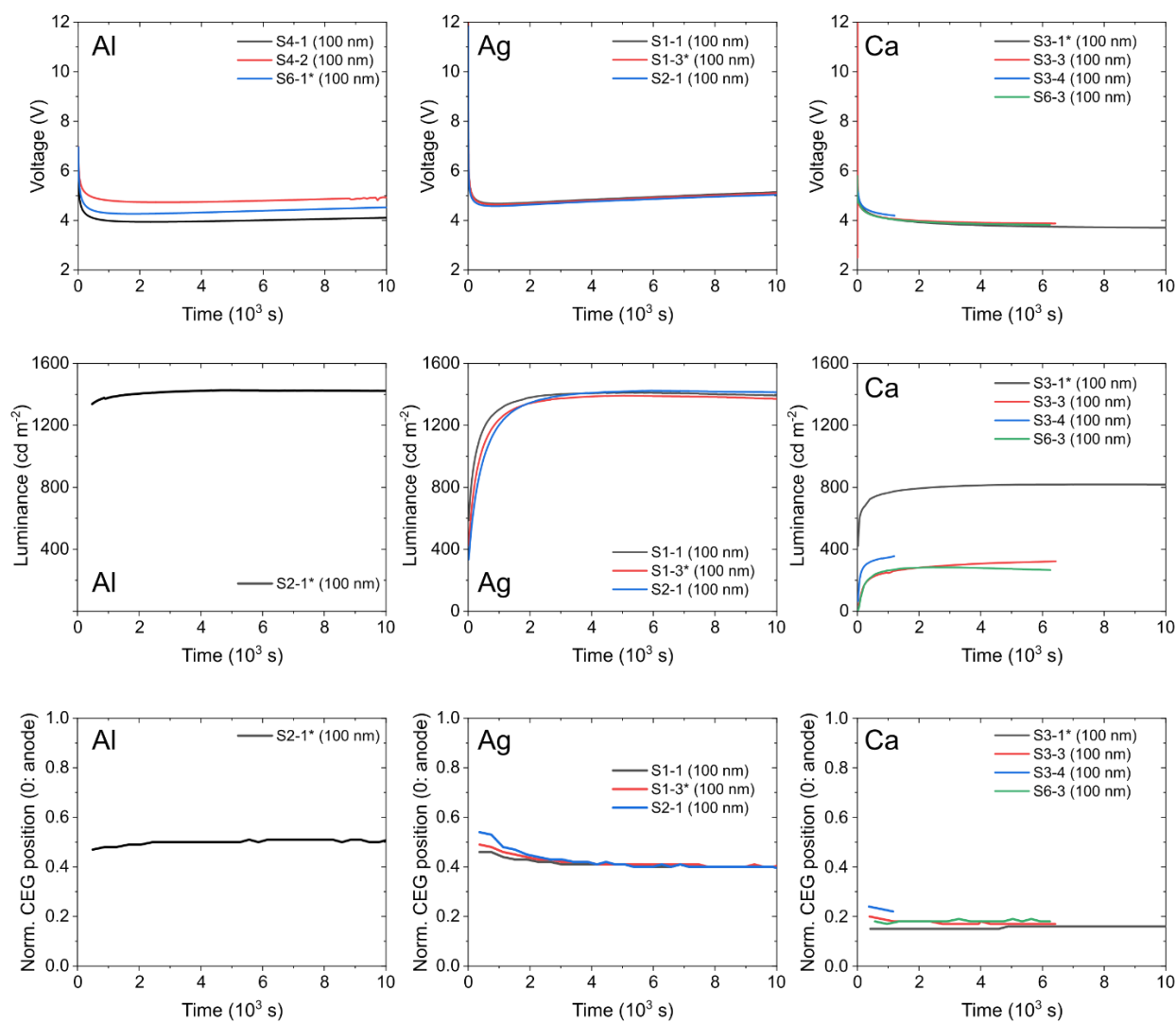

**Figure S3.** All experimental data for devices with $d_{AM} = 100$ nm.

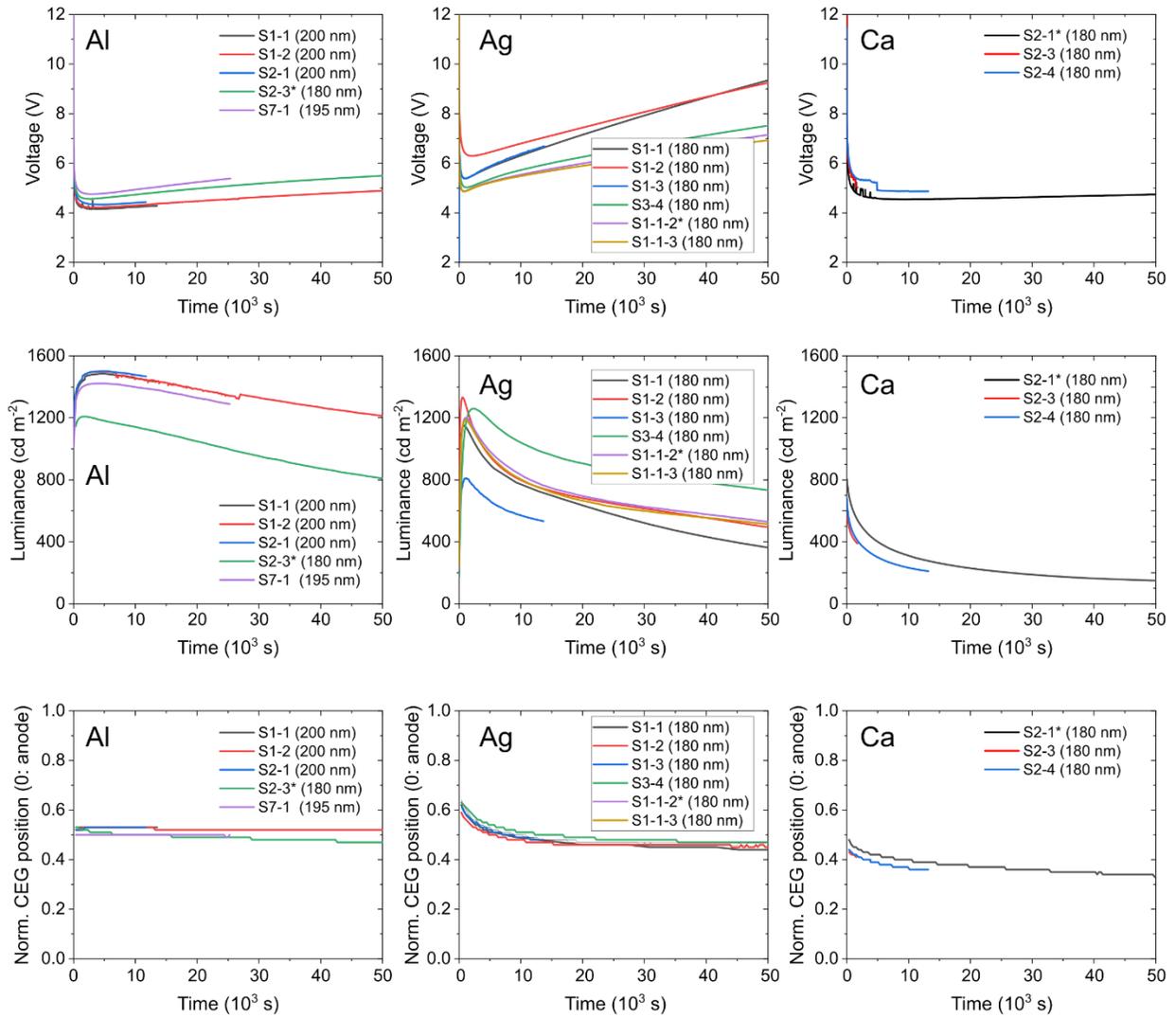

**Figure S4.** All experimental data for devices with $d_{AM} \approx 180$ nm.

The jumps in the calculated CEG position for thick devices, Figure S5, are an artifact of using a delta-distributed exciton generation profile $G(x)$ to reduce computation time for initial screening. For the data set selected for the main manuscript, the width of $G(x)$ was set as a parameter.

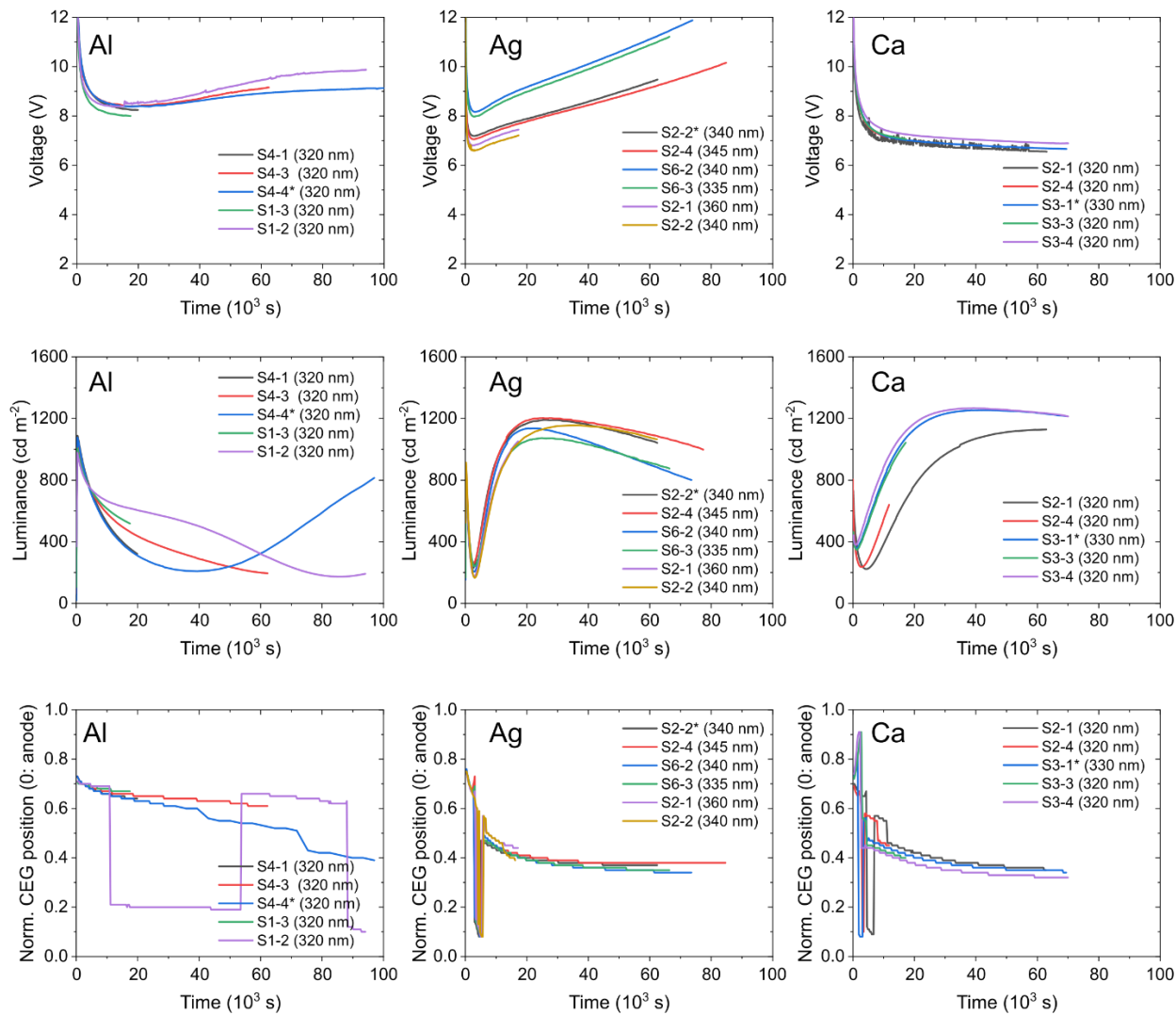

**Figure S5.** All experimental data for devices with $d_{AM} \approx 330$ nm.